\documentstyle[]{article}
\catcode`\@=11
\def\marginnote#1{}

\newcount\hour
\newcount\minute
\newtoks\amorpm
\hour=\time\divide\hour by60
\minute=\time{\multiply\hour by60 \global\advance\minute by-\hour}
\edef\standardtime{{\ifnum\hour<12 \global\amorpm={am}%
        \else\global\amorpm={pm}\advance\hour by-12 \fi
        \ifnum\hour=0 \hour=12 \fi
        \number\hour:\ifnum\minute<10 0\fi\number\minute\the\amorpm}}
\edef\militarytime{\number\hour:\ifnum\minute<10 0\fi\number\minute}

\def\draftlabel#1{{\@bsphack\if@filesw {\let\thepage\relax
   \xdef\@gtempa{\write\@auxout{\string
      \newlabel{#1}{{\@currentlabel}{\thepage}}}}}\@gtempa
   \if@nobreak \ifvmode\nobreak\fi\fi\fi\@esphack}
        \gdef\@eqnlabel{#1}}
\def\@eqnlabel{}
\def\@vacuum{}
\def\draftmarginnote#1{\marginpar{\raggedright\scriptsize\tt#1}}

\def\draft{\oddsidemargin -.5truein
        \def\@oddfoot{\sl preliminary draft \hfil
        \rm\thepage\hfil\sl\today\quad\militarytime}
        \let\@evenfoot\@oddfoot \overfullrule 3pt
        \let\label=\draftlabel
        \let\marginnote=\draftmarginnote
   \def\@eqnnum{(\theequation)\rlap{\kern\marginparsep\tt\@eqnlabel}%
\global\let\@eqnlabel\@vacuum}  }

\font\teneuf=eufm10  scaled  1200
\font\seveneuf=eufm7 scaled  1200
\font\fiveeuf=eufm5  scaled  1200

\newfam\euffam
\textfont\euffam=\teneuf  \scriptfont\euffam=\seveneuf
  \scriptscriptfont\euffam=\fiveeuf

\def\hexnumber@#1{\ifnum#1<10 \number#1\else
 \ifnum#1=10 A\else\ifnum#1=11 B\else\ifnum#1=12 C\else
 \ifnum#1=13 D\else\ifnum#1=14 E\else\ifnum#1=15 F\fi\fi\fi\fi\fi\fi\fi}

\def\got{\ifmmode\let\next\got@\else
 \def\next{\errmessage{Use \string\got\space only in math mode}}\fi\next}
\def\got@#1{{\got@@{#1}}}
\def\got@@#1{\fam\euffam#1}

\newfont{\lgot}{eufm10 scaled 2000}%

\newfont{\Bbb}{msbm10 scaled 1\@ptsize00}

\newcommand{\ZZ}{\mbox{\Bbb Z}}
\newfont{\Bbbb}{msbm7 scaled 1\@ptsize00}

\font\teneufm=cmmib10
\font\seveneufm=cmmib7
\font\fiveeufm=cmmib5
\def\bfit#1{{\textfont1=\teneufm\scriptfont1=\seveneufm
\scriptscriptfont1=\fiveeufm
\mathchoice{\hbox{$\displaystyle#1$}}{\hbox{$\textstyle#1$}}
{\hbox{$\scriptstyle#1$}}{\hbox{$\scriptscriptstyle#1$}}}}
\font\sevenmsa=msam6
\newfam\msafam
\textfont\msafam=\sevenmsa
\def\hexnumber@#1{\ifnum#1<10 \number#1\else
\ifnum#1=10 A\else\ifnum#1=11 B\else\ifnum#1=12 C\else
\ifnum#1=13 D\else\ifnum#1=14 E\else\ifnum#1=15 F\fi\fi\fi\fi\fi\fi\fi}
\def\msa@{\hexnumber@\msafam}
\mathchardef\blacktriangleright="3\msa@49
\mathchardef\blacktriangleleft="3\msa@4A

\newdimen\linethick  \linethick=0.4pt
\newdimen\hboxitspace    \hboxitspace=5pt
\newdimen\vboxitspace    \vboxitspace=5pt

\def\fr#1{%
\beq\new
\vcenter{
\hrule height\linethick
           \hbox{\vrule width\linethick
                 \kern\hboxitspace
                 \vbox{\kern\vboxitspace
                       \hbox{$\begin{array}{c}\displaystyle#1
          \end{array}$}%
                       \kern\vboxitspace}%
                 \kern\hboxitspace
                 \vrule width\linethick}%
           \hrule height\linethick}%
\eeq}

\renewcommand{\theequation}{\arabic{equation}}
\newcommand{\l@qq}[2]{\addvspace{2em}
 \hbox to\textwidth{\hspace{1em}\bf #1 \dotfill #2}}

\def\appname{Appendix}
\newcounter{app}
\def\theapp{\Alph{app}}
\def\app{\par
   \addvspace{4ex}
   \@afterindentfalse
  \secdef\@app\@dapp}
\def\@app[#1]#2{\ifnum \c@secnumdepth >\m@ne
        \refstepcounter{app}
        \addcontentsline{toc}{app}{\theapp
        \hspace{1em}#1}\else
      \addcontentsline{toc}{app}{ #1}\fi
   {\parindent \z@ \raggedright
    \Large \bf \appname~\theapp .
   \Large  \bf \hspace{1em}    #2}\nobreak
   \vskip 4ex   \noindent
\setcounter{equation}{0}
\def\theequation{\Alph{app}.\arabic{equation}}}
\def\@dapp#1{%
{\parindent \z@ \raggedright  \bf #1}\par\nobreak}
\def\l@app#1#2{\addpenalty{\@secpenalty}%
   \addvspace{1em plus\p@}%
   \begingroup
   \@tempdima 3em
     \parindent \z@ \rightskip \@pnumwidth
     \parfillskip -\@pnumwidth
     { \bf
     \leavevmode
     #1\hfil \hbox to\@pnumwidth{\hss #2}}\par
     \nobreak
   \endgroup}
\newcounter{sapp}[app]
\def\thesapp{\Alph{app}.\arabic{sapp}}
\def\sapp{\par
   \@afterindentfalse
  \secdef\@sapp\@dsapp}
\def\@sapp[#1]#2{\ifnum \c@secnumdepth >\m@ne
        \refstepcounter{sapp}
        \addcontentsline{toc}{sapp}{\thesapp
        \hspace{1em}#1}\else
      \addcontentsline{toc}{sapp}{ #1}\fi
   {\parindent \z@ \raggedright
    \large \bf \thesapp
   \large  \bf \hspace{1em}    #2}\nobreak
   \vskip 4ex   \noindent
\def\theequation{\Alph{app}.\arabic{equation}}}
\def\@dsapp#1{%
{\parindent \z@ \raggedright  \bf #1}\par\nobreak}
\def\l@sapp#1#2{\addpenalty{\@secpenalty}%
   \begingroup
   \@tempdima 3em
     \parindent \z@ \rightskip \@pnumwidth
     \parfillskip -\@pnumwidth
     { \hspace{1em}
     \leavevmode
     #1 \hfil \dotfill \hbox to\@pnumwidth{\hss #2}}\par \nobreak
     \endgroup}
\def\titlepage{\@restonecolfalse\if@twocolumn\@restonecoltrue\onecolumn
     \else \newpage \fi \thispagestyle{empty}\c@page\z@
}

\def\endtitlepage{\if@restonecol\twocolumn \else  \fi
        \def\thefootnote{\arabic{footnote}}
        \setcounter{footnote}{0}}  
\relax

\makeatletter
\newdimen\normalarrayskip              
\newdimen\minarrayskip                 
\normalarrayskip\baselineskip
\minarrayskip\jot
\newif\ifold             \oldtrue            \def\new{\oldfalse}
\def\arraymode{\ifold\relax\else\displaystyle\fi} 
\def\eqnumphantom{\phantom{(\theequation)}}     
\def\@arrayskip{\ifold\baselineskip\z@\lineskip\z@
     \else
     \baselineskip\minarrayskip\lineskip1\baselineskip\fi}

\def\@arrayclassz{\ifcase \@lastchclass \@acolampacol \or
\@ampacol \or \or \or \@addamp \or
   \@acolampacol \or \@firstampfalse \@acol \fi
\edef\@preamble{\@preamble
  \ifcase \@chnum
     \hfil$\relax\arraymode\@sharp$\hfil
     \or $\relax\arraymode\@sharp$\hfil
     \or \hfil$\relax\arraymode\@sharp$\fi}}

\def\@array[#1]#2{\setbox\@arstrutbox=\hbox{\vrule
     height\arraystretch \ht\strutbox
     depth\arraystretch \dp\strutbox
     width\z@}\@mkpream{#2}\edef\@preamble{\halign \noexpand\@halignto
\bgroup \tabskip\z@ \@arstrut \@preamble \tabskip\z@ \cr}%
\let\@startpbox\@@startpbox \let\@endpbox\@@endpbox
  \if #1t\vtop \else \if#1b\vbox \else \vcenter \fi\fi
  \bgroup \let\par\relax
  \let\@sharp##\let\protect\relax
  \@arrayskip\@preamble}
\def\eqnarray{\stepcounter{equation}%
              \let\@currentlabel=\theequation
              \global\@eqnswtrue
              \global\@eqcnt\z@
              \tabskip\@centering
              \let\\=\@eqncr
              $$%
 \halign to \displaywidth\bgroup
    \eqnumphantom\@eqnsel\hskip\@centering
    $\displaystyle \tabskip\z@ {##}$%
    &\global\@eqcnt\@ne \hskip 2\arraycolsep
         $\displaystyle\arraymode{##}$\hfil
    &\global\@eqcnt\tw@ \hskip 2\arraycolsep
         $\displaystyle\tabskip\z@{##}$\hfil
         \tabskip\@centering
    &{##}\tabskip\z@\cr}
\makeatother

\def\bea{\begin{eqnarray}}
\def\eea{\end{eqnarray}}

\def\beq{\begin{equation}}
\def\eeq{\end{equation}}
\def\be{\beq\new\begin{array}{c}}
\def\ee{\end{array}\eeq}
\def\2{{1\over 2}}

\def\balpha{{\bfit\alpha}}

\def\d{\partial}

\def\<{\langle}
\def\>{\rangle}
\def\ov{\overline}

\def\sem{\raise1pt\hbox{$\scriptscriptstyle >\!$}\:\!\!\tl}
\def\dr{\raise1pt\hbox{$\scriptscriptstyle >\!$}\!\!\btl}
\def\c{\epsilon}
\def\ep{\epsilon}
\def\f{1\over }
\def\ov{\overline}
\begin{document}
\thispagestyle{empty}
\baselineskip=12pt

\setcounter{footnote}0
\begin{center}
\hfill ITEP/TH-53/96\\
\hfill FIAN/TD-18/96\\
\hfill hep-th/9612094\\
\vspace{0.3in}
{\Large\bf Different Aspects of Relativistic Toda Chain}
\\[.3cm]
{\large S.Kharchev\footnote{ITEP, Bol.Cheremushkinskaya, 25, Moscow, 117 259,
Russia; kharchev@vxitep.itep.ru}}, {\large A.
Mironov\footnote{Theory Department,  P. N. Lebedev Physics
Institute, Leninsky prospect, 53, Moscow,
{}~117924, Russia and ITEP, Bol.Cheremushkinskaya, 25, Moscow, 117 259,
Russia; mironov@lpi.ac.ru,@heron.itep.ru}}, {\large A.
Zhedanov\footnote{Physics Department, Donetsk State University, Donetsk,
340 055, Ukraine and Donetsk Institute for Physics and Technology, Donetsk,
340 114, Ukraine; zhedanov@host.dipt.donetsk.ua}}
\end{center}
\bigskip
\centerline{\bf ABSTRACT}
\begin{quotation}
We demonstrate that the generalization of the relativistic Toda chain (RTC)
is a special reduction of two-dimensional Toda Lattice hierarchy (2DTL).
We also show that the RTC is gauge equivalent to the discrete
AKNS hierarchy and the unitary matrix model. Relativistic Toda molecule
hierarchy is also considered, along with the forced RTC. The simple approach
to the discrete RTC hierarchy based on Darboux-B\"acklund transformation is
proposed.
\end{quotation}


\bigskip

\noindent
{\Large\bf 1 Introduction}
\setcounter{footnote}{0}

\smallskip

\noindent
Since the paper of Ruijsenaars [1], where has been proposed, the
relativistic Toda chain (RTC) system was investigated in many papers
[2]-[4]. This system can be defined by
the equation:
\be\label{rt}
 \ddot q_{n} = \;\;(1+{\c}\dot q_{n})(1+{\c}\dot q_{n+1})
\frac{{\rm exp}(q_{n+1}-q_{n})}
{1+{\c^2}{\rm exp}(q_{n+1}-q_{n})} \;\;-\\
 - \;\;(1+{\c}\dot q_{n-1})(1+{\c}\dot q_{n})
\frac{{\rm exp}(q_{n}-q_{n-1})}
{1+{\c^2}{\rm exp}(q_{n}-q_{n-1})}
\ee
which transforms to the ordinary (non-relativistic) Toda chain (TC) in the
evident limit $\c\to 0$. The RTC is integrable, which was discussed
in different frameworks (see, for example, [2]-[4] and
references therein).
The RTC can be obtained as
a limit of the general Ruijsenaars system [1].

In this paper we are going to review different Lax representations of the
RTC, and to establish numerous realtions of it with many well-known integrable
systems like AKNS, unitary matrix model etc.
It is also shown that the RTC hierarchy can be embedded
to the 2DTL hierarchy [5].

Besides, we discuss the forced RTC hierarchy and its finite analog, the
relativistic Toda molecule. At the end of this short paper we
describe the simple approach to discrete evolutions
of the RTC which is based on the notion of the Darboux-B\"acklund
transformations and can be considered as a natural generalization of the
corresponding notion in the usual Toda chain theory.

This short paper is minded as a brief review.
For more details we refer the reader to our lengthy paper [6].

\bigskip

\noindent
{\Large\bf 2 Lax representation for RTC}

\smallskip

\noindent
Let us describe the Lax representation for the standard RTC
equation. The usual procedure to obtain integrable non-linear equations
consists of the two essential steps:

i)  To find appropriate spectral problem for the Baker-Akhiezer function(s).

ii) To define the proper evolution of this function with respect to
isospectral deformations.

{\bf Lax representation by three-term recurrent relation.}
In the theory of the usual Toda chain the first step implies the
dis\-cre\-tized ver\-sion of the Schr{\" o}\-din\-ger equa\-tion
(see [7], for example).
In order to get the relativistic extension of the Toda equations, one should
consider the following "unusual" spectral problem
\be \label{rec2}
\Phi_{n+1}(z) + a_{n}\Phi_{n}(z) =
z \{\Phi_{n}(z) + b_{n}\Phi_{n-1}(z)\}\;\; ,\;\;\;\;\;n\in\ZZ
\ee
representing a particular discrete Lax
operator acting on the Baker-Akhiezer function $\Phi_n(z)$.  This is a simple
three-term recurrent relation (similar to those for the
Toda and Volterra chains) but with
"unusual" spectral dependence.

As for the second step, one should note that there exist {\it two} distinct
integrable flows leading to the same equation (\ref{rt}). As we shall
see below, the spectral problem (\ref{rec2}) can be naturally incorporated
into the theory of two-dimensional Toda lattice (2DTL) which describes
the evolution with respect to two (infinite) sets of times
$\;\;(t_1, t_2,...)\;\;$,
$\;(t_{-1}, t_{-2},...)\;\;$
(positive and negative times, in accordance with [5]).
Here we describe the two particular flows (at the
moment, we deal with them "by hands", i.e. introducing the
corresponding Lax pairs
by a guess) which lead to the RTC equations (\ref{rt}).
The most simple evolution equation is
that with respect to the first {\it negative}
time and has the form
\be \label{todaev}
\frac{\d \Phi_{n}}{\d t_{-1}} =
R_{n}\Phi_{n-1}
\ee
with some (yet unknown) $R_{n}$.

The compatibility condition determines $R_{n}$ in terms of $a_{n}$ and
$b_{n}\ $ $R_{n} =\frac{b_{n}}{a_{n}}$
and leads to the following equations of motion:
\be \label{a-eq}
\frac{\d a_{n}}{\d t_{-1}} = \frac{b_{n}}{a_{n-1}} -
\frac{b_{n+1}}{a_{n+1}};
\ \ \ \ \
\frac{\d b_{n}}{\d t_{-1}} =
b_{n}\left(\frac{1}{a_{n-1}} - \frac{1}{a_{n}}\right)
\ee
In order to get (\ref{rt}), we should identify
\be\label{a-rep}
a_{n} = \exp(-\c p_{n});
\ \ \ \ \
b_{n} = -\c^{2}\exp(q_{n}-q_{n-1}-\c p_{n})
\ee
Note that in this parameterization the "Hamiltonian" $R_{n}$
depends only on coordinates $q_n$'s: $R_{n} = - \c^{2}\exp(q_{n}-q_{n-1})$.

Performing the proper rescaling of time in (\ref{a-eq})
we reach the RTC equation (\ref{rt}).

As we noted already, the evolution (\ref{todaev}), which leads to the RTC
equations is not the unique one. The other possible choice leading to the same
equations is
\be \label{todaev2}
\frac{\d \Phi_{n}}{\d t_{1}} = - b_{n}(\Phi_{n}-z\Phi_{n-1})
\ee
The compatibility condition of (\ref{rec2}) and (\ref{todaev2}) gives the
equations
\be \label{kuz1}
\frac{\d a_{n}}{\d t_{1}} = - a_{n}(b_{n+1} -b_{n})
\ \ \ \ \
\frac{\d b_{n}}{\d t_{1}} = -
b_{n}(b_{n+1} - b_{n-1} + a_{n-1} - a_{n})
\ee
This leads to the same RTC equation (\ref{rt}).

{\bf $2\times 2$ matrix Lax representation.}
The same RTC equation can be obtained from the matrix Lax operator
depending on the spectral parameter [3] (generalizing the Lax
operator for the TC [7]). Then the RTC arises as the compatibility
condition for the following $2\times 2$ matrix equations:
\be
L^{\scriptscriptstyle\rm (S)}_n \psi_n = \psi_{n+1}\;\;\; , \;\;\;
        \frac{\d \psi_{n}}{\d t} = A_{n}\psi_{n}
\ee
where
\be \label{l-op}
L^{\scriptscriptstyle\rm (S)}_{n} = \left(
\begin{array}{cc}
\zeta \exp(\c p_{n}) - \zeta^{-1} &  \c \exp(q_{n})\\
-\c \exp(-q_{n} + \c p_{n})             &  0
\end{array} \right) \;\;\; ; \hspace{5mm}
\psi_{n}=\left(
\begin{array}{c}
\psi^{(1)}_{n}\\
\psi^{(2)}_{n}
\end{array} \right)
\ee
\be \label{m-op}
A_{n} = \left(
\begin{array}{cc}
\c^{2} \exp(q_{n}-q_{n-1} + \c p_{n-1})    &  -\c\zeta^{-1}\exp(q_{n})\\
\c\zeta^{-1}\exp(-q_{n-1} +\c p_{n-1})  & 1 - \; \zeta^{-2}+\c^{2}
\end{array} \right)
\ee
One can easily reduce these equations to the system (\ref{rec2}) and
(\ref{todaev}).

To conclude this section, we remark that $\;L$-operator (\ref{l-op}),
which determines the RTC is not unique; moreover, it is not the simplest
one. Indeed, we shall see that there exists the whole
family of the gauge equivalent operators, which contains more "natural"
ones and includes, in particular, the well known operator generating
the AKNS hierarchy. From general point of view, the whole RTC hierarchy is
nothing but AKNS and vice versa.

\bigskip

\noindent
{\Large\bf 3 RTC and unitary matrix model, AKNS, etc.}

\smallskip

\noindent
Now we are going to describe the generalized RTC hierarchy as well as its
connection with some other integrable systems. We
start our investigation from the framework of
orthogonal polynomials

{\bf Unitary matrix model.}
It is well-known that the partition function $\;\tau_n\;$
of the unitary one-matrix model
can be presented as a product of norms of the biorthogonal polynomial
system [8]. Namely, let us introduce a scalar product of the form
\footnote{The signs of positive and negative times are defined in this
way to get the exact correspondence with the times introduced
in [5].}
\be\label{sp}
<A,B>=\oint {d\mu(z)\over 2\pi iz}
\exp\Bigl\{\sum_{m>0}(t_mz^m - t_{-m}z^{-m})\Bigr\}A(z)B(\frac{1}{z})
\ee
Let us define the system of polynomials biorthogonal with respect to
this scalar product
\be\label{orth-cond}
<\Phi_n,\Phi_k^{\star}>=h_n\delta_{nk}
\ee
Then, the partition function $\;\tau_n\;$of the unitary matrix model is
equal to the product of $h_n$'s:
\be\label{prod}
\tau_n = \prod_{k=0}^{n-1}h_k\;\;,\;\;\;\;\;\;\;\;\tau_0 \equiv 1
\ee
The polynomials are
normalized as follows:
\be\label{polyn}
\Phi_n(z)=z^n+\ldots+S_{n-1},\;\;\;\ \
\Phi_n^{\star}(z)=z^n+\ldots+ S^{\star}_{n-1},\;\;\;\; \ \
S_{-1}= S^{\star}_{-1}\equiv 1
\ee

These polynomials satisfy the following
recurrent relations:
\be \label{u-rec}
\Phi_{n+1}(z) = z\Phi_{n}(z)+ S_{n}z^{n} \Phi^{\star}_{n}(z^{-1})\\
\Phi^{\star}_{n+1}(z^{-1}) = z^{-1}\Phi^{\star}_{n}(z^{-1})+
S^{\star}_{n}z^{-n}\Phi_{n}(z)
\ee
and
\be\label{hS-rel}
{h_{n+1}\over h_n}=1-S_n S^{\star}_n
\ee

The above relations can be written in several equivalent forms. First, it
can be presented in the form analogous to (\ref{rec2}):
\be \label{q1}
\Phi_{n+1} - \frac{S_{n}}{S_{n-1}}\Phi_{n}
= z\left\{ \Phi_{n} - \frac{S_{n}}{S_{n-1}}
(1 - S_{n-1}S^{\star}_{n-1})\Phi_{n-1}\right\}
\ee
\be \label{q2}
\Phi^{\star}_{n+1} - \frac{
S^{\star}_{n}}{S^{\star}_{n-1}} \Phi^{\star}_{n} =
z^{-1}\left\{ \Phi^{\star}_{n} - \frac{
S^{\star}_{n}}{S^{\star}_{n-1}}(1 - S_{n-1} S^{\star}_{n-1})
\Phi^{\star}_{n-1}\right\}
\ee
{}From the first relation and using (\ref{rec2}) and (\ref{a-rep}),
one can immediately read off
\be\label{Shqp}
{S_n\over S_{n-1}}=-\exp(-\c p_n);
\ \ \ \ \
{h_n\over h_{n-1}}=-\c^2\exp(q_n-q_{n-1})
\ee
Thus, the orthogonality conditions (\ref{orth-cond}) lead exactly to
the spectral problem for the RTC. We should stress that equations
(\ref{q1}), (\ref{q2}) can be {\it derived} from the unitary matrix model.

Using the orthogonal conditions, it is also possible to obtain the equations
which describe the time dependence of $\;\Phi_n,\;\Phi^{\star}_n\;$.
Differentiating (\ref{orth-cond}) with respect to times
$\;t_1\;,\;\;t_{-1}\;$ gives the evolution equations:
\be\label{qt1}
\frac{\d \Phi_{n}}{\d t_{1}}=
\frac{S_{n}}{S_{n-1}}\frac{h_{n}}{h_{n-1}}
(\Phi_{n} -z \Phi_{n-1});
\ \ \ \ \
\frac{\d \Phi_{n}}{\d t_{-1}} =
\frac{h_{n}}{h_{n-1}}\Phi_{n-1}
\ee
\be \label{barqt1}
\frac{\d \Phi^{\star}_{n}}{\d t_{1}} =  -
\frac{h_{n}}{h_{n-1}}\Phi^{\star}_{n-1};
\ \ \ \ \
\frac{\d \Phi^{\star}_{n}}{\d t_{-1}}= -
\frac{S^{\star}_{n}}{S^{\star}_{n-1}}
\frac{h_{n}}{h_{n-1}}(\Phi^{\star}_{n} -z^{-1}\Phi^{\star}_{n-1})
\ee
(see general evolution equations with respect to higher flows below).
The compatibility conditions give the following nonlinear
evolution equations:
\be  \label{st1} \frac{\d S_{n}}{\d t_{1}} =
S_{n+1}\frac{h_{n+1}}{h_{n}};
\ \ \ \ \
\frac{\d S_{n}}{\d t_{-1}} = S_{n-1}\frac{h_{n+1}}{h_{n}}
\ee
\be  \label{barst1}
\frac{\d S^{\star}_{n}}{\d t_{1}} = - S^{\star}_{n-1}\frac{h_{n+1}}{h_{n}};
\ \ \ \ \
\frac{\d S^{\star}_{n}}{\d t_{-1}} = - S^{\star}_{n+1}\frac{h_{n+1}}{h_{n}}
\ee
As a consequence, in the polynomial case,
\be  \label{ht1}
\frac{\d h_{n}}{\d t_{1}} = - S_{n}S^{\star}_{n-1}h_n;
\ \ \ \ \
\frac{\d h_{n}}{\d t_{-1}} = S_{n-1}S^{\star}_{n}h_n
\ee
These are exactly relativistic Toda equations written in somewhat
different form.
Indeed, from (\ref{ht1}), (\ref{st1}), (\ref{barst1}) and (\ref{hS-rel})
one gets\footnote{The same equation holds for $t_{-1}$-flow.}
\be\label{h-eqs}
\frac{\d^2}{\d t_1^2}\log h_n = - \left(\frac{\d}{\d t_1}\log h_{n}\right)
\left(\frac{\d}{\d t_1}\log h_{n+1}\right)
\frac{{\displaystyle \frac{h_{n+1}}{h_n}}}
{\displaystyle {1-\frac{h_{n+1}}{h_n}}}\; + (n\to n-1)
\ee
On the other hand, the RTC is a particular case of the 2DTL hierarchy.
Indeed, let us introduce the key objects in the theory of integrable
systems - the $\tau$-functions,
which are defined through the relation $h_n=\tau_{n+1}/\tau_n$.
Then, with the help of (\ref{st1})-(\ref{ht1}), one can show that
the functions
$\;\tau_n\;$ satisfy the first equation of the 2DTL:
\be\label{todaeq}
\d_{t_1}\d_{t_{-1}}\log \tau_n = -\frac{\tau_{n+1}\tau_{n-1}}
{\tau^2_{n}}
\ee
Therefore, it is natural to assume that the higher flows generate the
whole set of non-linear equations of the 2DTL in spirit of [5].
This is indeed the case.

This completes the derivation of the RTC from the unitary matrix model.

{\bf RTC versus AKNS and "novel" hierarchies.}
Now let us demonstrate the correspondence between RTC and AKNS system.
We have already seen that the orthogonality conditions naturally lead
to the 2$\times$2 formulation of the problem generated by the unitary matrix
model:
\be\label{U1}
L^{\scriptscriptstyle\rm(U)}\left(
\begin{array}{c}
\Phi_n\\
\Phi^{\star}_n
\end{array}\right) = \left(
\begin{array}{c}
\Phi_{n+1}\\
\Phi^{\star}_{n+1}
\end{array}\right)\;\; ,
\;\;\;\;\;
L^{\scriptscriptstyle\rm(U)}=\left(
\begin{array}{cc}
    z             &  \;\;\;\; z^nS_n\\
z^{-n}S_n^{\star} &  \;\;\;\; z^{-1}
\end{array}\right)
\ee
\be\label{U2}
\frac{\d}{d t_1}\left(
\begin{array}{c}
\Phi_n\\
\Phi^{\star}_n
\end{array}\right)=\left(
\begin{array}{cc}
-S_nS_{n-1}^{\star} & \;\;\;z^nS_n\\
z^{1-n}S^{\star}_{n-1}  & \;\;\;-z
\end{array}\right)\left(
\begin{array}{c}
\Phi_n\\
\Phi^{\star}_n
\end{array}\right)
\ee
\be\label{U3}
\frac{\d}{d t_{-1}}\left(
\begin{array}{c}
\Phi_n\\
\Phi^{\star}_n
\end{array}\right)=\left(
\begin{array}{cc}
       z^{-1}             & \;\;\; -z^{n-1}S_{n-1}\\
-z^{-n}S^{\star}_{n}  & \;\;\; S_{n-1}S_{n}^{\star}
\end{array}\right)\left(
\begin{array}{c}
\Phi_n\\
\Phi^{\star}_n
\end{array}\right)
\ee
(Equations (\ref{U2}), (\ref{U3}) follow from (\ref{qt1})-(\ref{barqt1})
and the original spectral problem (\ref{u-rec})).
Put $
\Phi_{n} \equiv z^{n/2-1/4}F_{n},\ \
\Phi^{\star}_{n} \equiv z^{-n/2+1/4}F^{\star}_{n}$.
Then the spectral problem (\ref{U1}) can be rewritten in the matrix form
\be
L_{n}^{\scriptscriptstyle\rm (AKNS)}
{\cal F}_{n} = {\cal F}_{n+1} \;\; , \;\;\;\;\;
{\cal F}_{n} \equiv \left(
\begin{array}{l}
F_{n}\\
F^{\star}_{n}
\end{array} \right)
\ee
where
\be \label{akns}
L_{n}^{\scriptscriptstyle\rm (AKNS)} = \left(
\begin{array}{ll}
\zeta           &   S_{n}\\
S^{\star}_{n}   &   \zeta^{-1}
\end{array} \right) \;\;, \;\;\;\;\;\; \zeta \equiv z^{1/2}
\ee
This is the Lax operator for the discrete AKNS [9].
Obviously the evolution equations (\ref{U2}), (\ref{U3}) can be written
in terms of $F_{n},\;
F^{\star}_{n}$ as
\be
\frac{\d {\cal F}_{n}}{\d t_{1}} = A^{(1)}_{n}{\cal F}_{n}\;\; ,
\hspace{12mm} A^{(1)}_{n} = \left(
\begin{array}{ll}
- S_{n}S^{\star}_{n-1}        &   \zeta S_{n}\\
\zeta S^{\star}_{n-1}       &   -\zeta^{2}
\end{array} \right)
\ee
\be
\frac{\d {\cal F}_{n}}{\d t_{-1}} =\;-\;
A^{(-1)}_{n}{\cal F}_{n}\;\; ,
\hspace{12mm} A^{(-1)}_{n} = \left(
\begin{array}{ll}
-\zeta^{-2}              &   \zeta^{-1} S_{n-1}\\
\zeta^{-1} S^{\star}_{n}    &   - S_{n-1}S^{\star}_{n}
\end{array} \right)
\ee
Note that after introducing the trivial flow
\be
\frac{\d {\cal F}_{n}}{\d t_{0}} =
A^{(0)}_{n}{\cal F}_{n}\;\; ,
 \hspace{12mm} A^{(0)}_{n} = \left(
\begin{array}{ll}
1  &  0 \\
0  &  -1
\end{array} \right)
\ee
we get the difference non-linear Schr{\" o}dinger system (DNLS)
[9] (see also [7]) generated by the "mixed" flow
\be\label{T-dep}
\frac{\d {\cal F}_{n}}{\d T} \equiv
\left( \frac{\d}{\d t_{0}}  - \frac{\d}{\d t_{-1}}
- \frac{\d}{\d t_{1}}\right) {\cal F}_{n}
= (A^{(0)}_{n} + A^{(-1)}_{n} - A^{(1)}_{n}){\cal F}_{n}\equiv\\
\equiv\;\;\left(
\begin{array}{ll}
1+S_nS_{n-1}^{\star} -\zeta^{-2}\;\;\;\; & \zeta^{-1}S_{n-1} - \zeta S_n \\
\zeta^{-1}S_{n}^{\star} - \zeta S_{n-1}^{\star} \;\;\;\;&
-1 - S_{n-1}S_n^{\star} +\zeta^2
\end{array}\right)
\ee
Indeed, from the compatibility conditions for (\ref{akns}), (\ref{T-dep})
or, equivalently, just from (\ref{st1}) (along with the trivial
evolution $
\d_{t_0} S_n = 2S_n\;,\;
\d_{t_0} S^{\star}_n= -2S^{\star}_n\;$)
one gets the discrete version of the nonlinear Schr\"odinger equation:
\be
\frac{\d S_n}{\d T} = - (S_{n+1} - 2S_n + S_{n-1}) + S_nS_n^{\star}
\bigl( S_{n+1} + S_{n-1}\bigr)
\ee
Note also that the "novel" hierarchy of [10] is equivalent to the RTC
(and, therefore, to the AKNS hierarchy) as well. Namely, the Lax operator in
[10], i.e.
\be \label{new}
{\widehat L}_{n} = \left(
\begin{array}{cc}
z + u_{n}v_{n}    &  \;\;\;  u_{n}\\
     v_{n}        &   1
\end{array} \right) \;\; ; \;\;\;\;\;\;\;\;\;\;\;
{\widehat L}_n\left(
\begin{array}{l}
\phi_n^{(1)}\\
\phi_n^{(2)}
\end{array}\right) =   \left(
\begin{array}{l}
\phi_{n+1}^{(1)}\\
\phi_{n+1}^{(2)}
\end{array}\right)
\ee
defines the recurrent relation of the form (\ref{q1}):
\be \label{rag1}
\phi^{(1)}_{n+1} - \left( u_{n}v_{n} + \frac{u_{n}}{u_{n-1}}\right)
 \phi^{(1)}_{n} = z \left( \phi^{(1)}_{n} -
 \frac{u_{n}}{u_{n-1}}\phi^{(1)}_{n-1} \right)
\ee
thus revealing the connection with the RTC.
Comparing (\ref{q1}) and (\ref{rag1}) leads to the
identification $
u_{n} = S_{n}h_{n} \;\;\; ,\hspace{10mm}
v_{n} = \frac{S^{\star}_{n-1}}{h_{n}}$,
where $h_{n}$'s satisfy (\ref{hS-rel}). Moreover, from (\ref{new})
and (\ref{u-rec}) it is easy to see that $
\phi^{(1)}_n = \Phi_n;
\ \
\phi^{(2)}_n = \frac{1}{h_n}\Bigl(z^n\Phi^{\star}_n -
S^{\star}_{n-1}\Phi_n\Bigr)$
and, therefore, $\;{\widehat L}_n\;$ can be obtained from
$\;L_{n}^{\scriptscriptstyle\rm (AKNS)}\;$ by the discrete gauge
transformation:
\be
{\widehat L}_{n} =
U_{n+1}L_{n}^{\scriptscriptstyle\rm (AKNS)}U_{n}^{-1};
\ \ \ \ \ \
U_n = z^{n/2-1/4}\left(
\begin{array}{cc}
1\; & 0\\
-\frac{S_{n-1}^{\star}}{h_n} & \frac{z^{1/2}}{h_n}
\end{array}\right);\ \ \ \ z=\zeta^2
\ee
Evolution equations (\ref{st1})-
(\ref{ht1}) in terms of new variables $u_n,\ v_n$ have the form
\be
\begin{array}{ll}
\frac{\d u_n}{\d t_1} = u_{n+1} - u_n^2v_n \;\; , \;\;\;\;\;\;\;\; &
\frac{\d u_n}{\d t_{-1}} = \frac{u_{n-1}}{1+ u_{n-1}v_n}\\
\frac{\d v_n}{\d t_1} = - v_{n-1} + u_nv_n^2 \;\; , \;\;\;\;\;\;\;\; &
\frac{\d v_n}{\d t_{-1}} = - \frac{v_{n+1}}{1+ u_{n}v_{n+1}}
\end{array}
\ee
and easily reproduce the usual AKNS equations in the continuum
limit since
\be
\bigl(\d_{t_{0}} - \d_{t_{1}}-\d_{t_{-1}}\bigr)u_n  =
-\bigl(u_{n+1} -2u_n + u_{n-1}\bigr) +
\bigl(u^2_{n-1}+u^2_n\bigr)v_n + \ldots \\
\bigl(\d_{t_{0}} - \d_{t_{1}}-\d_{t_{-1}}\bigr)v_n  =
\bigl(v_{n+1} -2v_n + v_{n-1}\bigr) -
\bigl(v^2_{n}+v^2_{n+1}\bigr)u_n + \ldots
\ee
We conclude with
the remark that the operator $\;L_n^{\scriptscriptstyle\rm (S)}\;$
in (\ref{l-op}) is also gauge equivalent to
$\;L_{n}^{\scriptscriptstyle\rm (AKNS)}\;$ (see [6]).

{\bf Non-local Lax representation.}
There is another form of the
recurrent relations which is non-local (i.e. contains all the functions with
smaller indices) but instead expresses  $\Phi_n(z)$ through
themselves. This form is crucial for dealing
with the RTC as a particular reduction of the 2DTL.
Let us introduce the normalized functions
\be\label{PP}
{\cal P}_n(z) \equiv  \Phi_n(z)\ \;\;,\;\;\;\;\;\;
        {\cal P}^{\star}_n(z^{-1}) \equiv  {1\over h_n}
    \Phi^{\star}_n(z^{-1})
\ee
From (\ref{q1}), (\ref{q2}) one can show
that in the forced and fast-decreasing cases some proper solutions
satisfy the equations
\be\label{Lax2DTL}
z{\cal P}_n(z) =
{\cal P}_{n+1}(z) - S_nh_n\sum ^n_{k=-\infty} {S^{\star}_{k-1}\over h_k}
{\cal P}_k(z) \equiv  {\cal L}_{nk}{\cal P}_k(z)\\
z^{-1}{\cal P}^{\star}_n(z^{-1}) =
{h_{n+1}\over h_n} {\cal P}^{\star}_{n+1}(z^{-1}) - S^{\star}_n
\sum^n_{k=-\infty}S_{k-1} {\cal P}^{\star}_k(z^{-1}) \equiv
\ov{\cal L}_{kn} {\cal P}^{\star}_k(z^{-1})
\ee
This expression is correct for
general (non-polynomial) ${\cal P}_n$ and ${\cal P}^{\star}_n$ provided the
sums run over all integer $k$. In the polynomial case, the sums automatically
run over only non-negative $k$.
The last representation of the spectral problem will be useful
to determine the general evolution of the system. Indeed,
these relations manifestly describe the embedding of the RTC into the
2DTL [5,11], which is given essentially by {\it two} Lax operators
(${\cal L}$ and $\ov{\cal L}$).

{\bf RTC as reduction of 2DTL.}
In order to determine the whole set of the evolution equations, one can
use different tricks. For example, one can use embedding (\ref{Lax2DTL})
of the system into the 2DTL, making
use of the standard evolution of this latter [5].
Let us briefly
describe the formalism of the 2DTL following [5]. In their framework,
one introduces {\it two} different Baker-Akhiezer (BA) $\ZZ\times\ZZ$
matrices ${\cal W}$ and $\ov{\cal W}$.
These matrices satisfy the linear system:\\
\phantom{1cm}i) the matrix version of the spectral problem:
\be\label{OE2}
{\cal L}{\cal W} = {\cal W} \Lambda\;\;,\ \ \
\ov{{\cal L}}\;\ov{\cal W} = \ov{\cal W}\Lambda ^{-1}
\ee
\phantom{1cm}ii) the matrix version of the evolution equations:
\be\label{OE3}
\begin{array}{lll}
\frac{\d {\cal W}}{\d t_{m}}=({\cal L}^{m})_{+}{\cal W}\;,\;\;\;&
\frac{\d \ov{{\cal W}}}{\d t_{m}}=
({\cal L}^{m})_{+}\ov{{\cal W}}&\\
\frac{\d {\cal W}}{\d t_{-m}}=(\ov{\cal L}^{\;m})_{-}{\cal W}\;,\;\;\;&
\frac{\d \ov{{\cal W}}}{\d t_{-m}}=(\ov{\cal L}^{\;m})_{-}\ov{{\cal W}}
\;,\;\;\;&m=1,2,...
\end{array}
\ee
where $\ZZ\times\ZZ$ matrices ${\cal L}$ and $\ov{\cal L}$ have
(by definition) the following structure:
\be \label{lax1}
{\cal L} = \sum_{i\leq 1}{\rm diag}[b_{i}(s)]\Lambda^i \;\; ;\;\;\;
	b_{1}(s)=1\\
\ov{\cal L} = \sum_{i\geq -1}{\rm diag}[c_{i}(s)]\Lambda^i \;\; ;\;\;\;
	c_{-1}(s)\neq 0
\ee
Here ${\rm diag}[b_{i}(s)]$ denotes an infinite diagonal matrix
$\;{\rm diag}(\ldots \;b_{i}(-1),\; b_{i}(0), \;b_{i}(1),\;\ldots$) ;$\;\;\;$
$\Lambda$ is the shift matrix with the elements $\Lambda_{nk}
\equiv \delta_{n,k-1}$ and for arbitrary infinite matrix
$A=\sum_{i\in\ZZ}{\rm diag}[a_{i}(s)]\Lambda^{i}$ we set
\be
(A)_{+}\equiv\sum_{i\geq 0}{\rm diag}[a_{i}(s)]\Lambda^{i}\;\;,\;\;\;\;\;
(A)_{-}\equiv\sum_{i<0}{\rm diag}[a_{i}(s)]\Lambda^{i}
\ee
i.e. $(A)_{+}$ is the upper triangular part of the matrix $A$ (including
the main diagonal) while $(A)_{-}$ is strictly the lower triangular part.

Note that (\ref{lax1}) can be written in components as
\be\label{lax2}
{\cal L}_{nk}=\delta_{n+1,k} +
b_{k-n}(n)\theta (n-k)\;,\;\;\;
\ov{\cal L}_{nk}=c_{-1}(n)\delta_{n-1,k} +
c_{k-n}(n)\theta (k-n) \;\; ; \;\;\;\;n,k\in\ZZ
\ee
The compatibility conditions
imposed on (\ref{OE2}),(\ref{OE3}) give rise
to the infinite set (hierarchy) of
nonlinear equations for the operators ${\cal L}$,
$\ov{\cal L}$ or, equivalently, for the coefficients $b_{m}(n)$, $c_{m}(n)$.
This is what is called 2DTL hierarchy.

{}From (\ref{Lax2DTL}), one gets two matrices
\be\label{RTC1}
{\cal L}_{nk} = \delta_{n+1,k} -\frac{h_n}{h_k}S_{n}S^{\star}_{k-1}
\theta(n-k)\;\;,\;\;\;\;\;k, n \in \ZZ
\ee
\be\label{RTC2}
\ov{\cal L}_{nk} = \frac{h_n}{h_{n-1}}\delta_{n-1,k}
-S_{n-1}S^{\star}_{k}\theta(k-n)\;\;,\;\;\;\;\;k, n \in \ZZ
\ee
which have exactly the form (\ref{lax2}). Now using the technique developed
in [5], one can get the whole evolution of the RTC hierarchy.
However, it can be also easily obtained in the framework of the orthogonal
polynomials.

{\bf Evolution and orthogonal polynomials.}
Differentiating the orthogonality conditions with
respect to arbitrary times, one can obtain with the help of (\ref{Lax2DTL})
the evolution of polynomials  ${\cal P}_n$ and  ${\cal P}^{\star}_n$
[6]:
\be\label{OE1}
{\partial {\cal P}_n\over \partial t_m} =
- [({\cal L}^m)_-]_{nk}{\cal P}_k;
\ \ \ \ \
{\partial {\cal P}_n\over \partial t_{-m}} = [(\ov{\cal L}^{\;m})_-]_{nk}
{\cal P}_k
\\
{\partial {\cal P}^{\star}_n\over \partial t_m} = -
[({\cal L}^m)_+]_{kn}{\cal P}^{\star}_k;
\ \ \ \ \
{\partial {\cal P}^{\star}_n\over \partial t_{-m}} =
[(\ov{\cal L}^{\;m})_+]_{kn}{\cal P}^{\star}_k\\
{\partial h_n\over \partial t_m} = ({\cal L}^m)_{nn}h_n\ \;\;,\;\;\;\;\;\;
{\partial h_n\over \partial t_{-m}} = - (\ov{\cal L}^{\;m})_{nn}h_n
\ee

\bigskip

\noindent
{\Large\bf 4 Forced RTC hierarchy}

\smallskip

\noindent
{\bf RTC-reduction of 2DTL.}
Let us formulate in some invariant terms what reduction of
the 2DTL
corresponds to the RTC hierarchy. Return again to the
Lax representation (\ref{Lax2DTL}) embedding the RTC into the 2DTL.
Using (\ref{hS-rel}), one can easily prove the following identities
\be\label{zm}
\sum_{k=n}^{N}{S_{k-1}S^{\star}_{k-1}\over h_k}={\f h_N}-{\f h_{N-1}};
\ \ \ \ \
\sum_{k=n}^N S_kS^{\star}_kh_k=h_n-h_{N+1}
\ee
Because of these identities, the matrices ${\cal L}$ and ${\bar {\cal L}}^{T}$
have zero modes $\sim S_{k-1}$ and $S^{\star}_{k-1}/h_k$ respectively.
Therefore, one could naively expect that they are not invertible and get
(using (\ref{zm})) that
\be\label{LM=1}
({\cal L}\bar {\cal L})_{nk}=\delta_{nk}-{S_nS^{\star}_kh_n\over
h_{-\infty}};
\ \ \ \ \
(\bar {\cal L}{\cal
L})_{nk}=\delta_{nk}-{S_{n-1}S^{\star}_{k-1}h_{\infty}\over h_k}
\ee
Since
the reduction is to be described as an invariant condition imposed on ${\cal
L}$ and $\bar {\cal L}$, these formulas might serve as a starting point to
describe the reduction of the 2DTL to the RTC hierarchy only if their r.h.s.
does not depend on the dynamical variables. It seems not to be the case.

However, these formulas require some careful treatment.
Indeed, the formulation of the 2DTL in terms of infinite-dimensional matrices,
although being correct as a formal construction requires some accuracy if one
wants to work with the genuine matrices since the products of the
infinite-dimensional matrices should be properly defined. In fact, this product
exists for the "band" matrices, i.e. those with only a finite number of
the non-zero diagonals, and in some other more complicated cases (of special
divergency conditions). One can easily see from  (\ref{Lax2DTL}) that the RTC
Lax operators do not belong to this class. Therefore, equations (\ref{LM=1})
just do not make sense in this case (this is why the interpretation of the
general RTC hierarchy in invariant (say, Grassmannian) terms
is a little bit complicated).

However, in the case of forced hierarchy, some of the indicated problems are
removed since one needs to multiply only
quarter-infinite matrices and, say, the product ${\cal L}\bar{\cal L}$
always exists. Certainly the inverse order of the multipliers is still
impossible. Therefore, only the first formula in (\ref{LM=1})
becomes well-defined acquiring the form
\be\label{forcedLM=1}
({\cal L}\bar {\cal L})_{nk}=\delta_{nk}
\ee
This formula can be already
taken as a definition of the RTC-reduction of the 2DTL
in the forced case as it does not depend on dynamical variables.
Now we will show how this definition is reflected
in different formulations of the 2DTL.

{\bf Determinant representation.}
One can show (see [6]) that $\tau$-function
of the forced hierarchy $\tau_n=0,\ n<0$ has the
following determinant representation [12]
\beq\label{detrep}
\tau _n(t) = \left.
\det \left[\partial ^i_{t_1}(-\partial _{t_{-1}})^j\int _\gamma
A(z,w)\exp\left\{\sum_{m>0} (t_mz^m- t_{-m}w^{-m})\right\}dzdw\right]
\right|
_{i,j=0,...,n-1}
\eeq
where the matrix $A$ has the form
$A(z,w)={\mu'(z)\over 2\pi iz}\delta (z-w^{-1})$ [6].

Indeed, let us rewrite the orthogonality relation (\ref{orth-cond})
in the matrix form. We define matrices $D$ and $D^{\star}$
with the matrix elements determined as the coefficients of the polynomials
$\Phi_n(z)$ and $\Phi_n^{\star}(z)$
\be
\Phi_i(z)\equiv \sum_jD_{ij}z^{j-1},\ \ \ \Phi^{\star}_i(z)\equiv
\sum_j D^{\star}_{ij}z^{j-1}
\ee
Then, (\ref{orth-cond}) looks like
\be\label{oco}
D\cdot C\cdot D^{\star T}=H
\ee
where superscript $T$ means transponed matrix and $H$ denotes the diagonal
matrix with the entries $C_{ii}=h_{i-1}$
and $C$ is the moment matrix with the matrix
elements
\be
C_{ij}\equiv \int_{\gamma} {d\mu(z)\over 2\pi iz} z^{i-j}
\exp\left\{\sum_{m>0}(t_mz^m-t_{-m}z^{-m})
\right\}
\ee
Let us note that $D$ ($D^{\star T}$) is
the upper (lower) triangle matrix with the units on the diagonal (because of
(\ref{polyn})). This representation is nothing but the
Riemann-Hilbert problem for the forced hierarchy.
Now taking the determinant of the both sides of (\ref{oco}),
one gets
\be
\det_{n\times n} C_{ij}=\prod_{k=0}^{n-1} h_k =\tau_n
\ee
due to formula (\ref{prod}). The remaining last step is to observe that
\be\label{moma}
C_{ij}=\partial ^i_{t_1}(-\partial _{t_{-1}})^j\int _\gamma
{d\mu(z)\over 2\pi iz}
\exp\left\{\sum_{m>0}(t_mz^m-t_{-m}z^{-m})\right\}=
\partial ^i_{t_1}(-\partial _{t_{-1}})^j C_{11}
\ee
i.e.
\be\label{taumoma}
\tau_n(t)=\det_{n\times n}\left[
\partial ^i_{t_1}(-\partial _{t_{-1}})^j\int _\gamma
{d\mu(z)\over 2\pi iz}
\exp\left\{\sum_{m>0}(t_mz^m-t_{-m}z^{-m})\right\}
\right]
\ee
This expression coincides with (\ref{detrep}).
One can also remark that the moment matrix $C_{ij}$ is Toeplitz
matrix. This proves from the different
approach that the RTC-reduction is defined by the Toeplitz matrices.

\bigskip

\noindent
{\Large\bf 5 Relativistic Toda molecule}

\smallskip

\noindent
{\bf General properties.}
Let us consider further restrictions on the RTC which allows one
to consider the {\it both} products in (\ref{LM=1}). Namely, in addition to
the condition $\tau_n=0,\ n<0$ picking up forced hierarchy, we impose the
following constraint
\be\label{molecule}
\tau_n=0,\ \ \ n>N
\ee
for some $N$.
This system should be called $N-1$-particle
relativistic Toda molecule, by analogy
with the non-relativistic case and is nothing but RTC-reduction of the
two-dimensional Toda molecule [13,14]\footnote{Sometimes
the Toda molecule
is called non-periodic Toda [15]. It is an
immediate generalization
of the Liouville system.}.

$sl(N)$ Toda can be described by the kernel $A(z,w)$
\be\label{kernel}
A(z,w)=\sum_k^N f^{(k)}(z)g^{(k)}(w)
\ee
where $f^{(k)}(z)$ and $g^{(k)}(z)$ are arbitrary functions.
{}From this description, one can immediately read off the corresponding
determinant representation (\ref{detrep}).

Indeed, equation (\ref{todaeq}) and condition (\ref{molecule}) implies
that $\log\tau_0$ and $\log\tau_N$ satisfy the free wave equation
$\partial_{ t_1}\partial_{ t_{-1}}\log \tau_0=
 \partial_{ t_1}\partial_{ t_{-1}}\log \tau_{N}=0$.
Since the relative normalization of $\tau_n$'s is not fixed, we are free
to choose $\tau_0=1$. Then, $\label{mol}\tau_{0}(t)=1,\ \
\tau_{N}(t) = \chi(t_1)\bar \chi( t_{-1})$,
where $\chi(t_1)$ and $\bar \chi(t_{-1})$ are arbitrary functions. 2DTL with
boundary conditions (\ref{mol}) was considered in [13]. The solution
to (\ref{todaeq}) in this case is given by [14]:
\be\label{C1}
\tau_{n}(t)\;=\;
	\det\;\partial_{t_1}^{i-1} (-\partial_{t_{-1}})^{j-1}\tau_{1}(t)
\ee
with
\be
\tau_{1}(t) = \sum_{k=1}^{N} a^{(k)}(t)\bar a^{(k)}(t_{-1})
\ee
where functions $a^{(k)}(t)$ and $\bar a^{(k)}(t_{-1})$ satisfy
$$\det\partial^{i-1}_{t_1}a^{(k)}(t)=\chi(t),\ \ \ \ \
\det (-\partial_{t_{-1}})^{i-1}\bar a^{(k)}(t_{-1}) = \bar \chi(t_{-1})$$
This result coincides with that obtained by substituting into (\ref{detrep})
the kernel $A(z,w)$ of the form (\ref{kernel}).

{\bf Lax representation.}
In all our previous considerations, we dealt with
infinite\--di\-men\-si\-o\-nal matrices.  Let us note that the Toda molecule
can be effectively treated in terms of $N\times N$ matrices like the forced
case could be described by the quarter-infinite matrices. This allows one to
deal with the {\it both} identities (\ref{LM=1}) since all the products of
{\it finite} matrices are well-defined.

To see this, one can just look at the recurrent relation (\ref{Lax2DTL}) and
observe that there exists
the finite-dimensional subsystem of ($N$) polynomials which is decoupled from
the whole system. The recurrent relation for these polynomials can be
considered
as the finite-matrix Lax operator (which still
does not depend on the spectral parameter, in contrast to (\ref{l-op})).
Indeed, from (\ref{Lax2DTL}) and
condition (\ref{molecule}), i.e. $h_n/h_{n-1}=0$ as $n\ge N$ (the Toda
molecule conditions in terms of $S$-variables read as $S_n=S^{\star}_n=1$
for $n>N-2$ or $n>0$),
one can see that
\be
z{\cal P}_N(z)={\cal P}_{N+1}(z)-{\cal P}_N(z),\ \ \
z{\cal P}_{N+1}(z)={\cal P}_{N+2}(z)-{\cal P}_{N+1}(z)\ \ \ \hbox{etc.}
\ee
i.e. all the polynomials ${\cal P}_n$ with $n>N$ are trivially expressed
through ${\cal P}_N$. Therefore, the system can be effectively described by
the dynamics of only some first polynomials (i.e. has really finite number of
degrees of freedom). Certainly, all the same is correct for the
star-polynomials ${\cal P}_n^{\star}$ although, in this case, it would be
better
to use the original non-singularly normalized polynomials $\Phi_n^{\star}$.

Now let us look at the corresponding Lax operators (\ref{RTC1})-(\ref{RTC2}).
They are getting quite trivial everywhere but in the left upper corner
of the size $N\times N$ [6].
Therefore, one can
restrict himself to the system of $N$ polynomials ${\cal P}_n$, $n=0,1,\ldots,
N-1$ and the finite matrix Lax operators (of the size $N\times N$).

Now one needs only to check that this finite system still has the same
evolution equations (\ref{OE1}). It turns out to be the case only for the
first $N-1$ times. This is not so surprising, since, in the finite system
with $N-1$ degrees of freedom, only first $N-1$ time flows are independent.
Therefore, if looking at the finite matrix Lax operators, one gets the
dependent higher flows. On the other hand, if one embeds this finite system
into
the infinite 2DTL, one observes that the higher flows can be no longer
described inside this finite system. Just the finite
system is often called relativistic Toda molecule.

To simplify further notations, we introduce, instead of $S_n$, $S^{\star}_n$,
the new variables $s_n\equiv (-)^{n+1} S_n$, $s^{\star}_n\equiv (-)^{n+1}
S^{\star}_n$. Then, one can realize that the Lax operator
can be constructed as the product of simpler ones
$
{\cal L}={\cal L}_N{\cal L}_{N-1}\ldots{\cal L}_1
$,
where ${\cal L}_k$ is the unit matrix wherever but a $2\times 2$-block:
\be\label{L4}
{\cal L}_k\equiv
\left(\begin{array}{ccc}
1&\vdots&\\
\cdots&G_k&\cdots\\
&\vdots&1
\end{array}
\right)\ \ \ \ \ \ G_k\equiv\left(\begin{array}{cc}
s_k&1\\
s_ks^{\star}_k-1&s^{\star}_k
\end{array}\right)
\ee
Analogously, $\bar {\cal L}=\bar {\cal L}_1\ldots\bar {\cal L}_{N-1}\bar
{\cal L}_N$
with
\be
\bar {\cal L}_k\equiv
\left(\begin{array}{ccc}
1&\vdots&\\
\cdots&\bar G_k&\cdots\\
&\vdots&1
\end{array}
\right)\ \ \ \ \ \ \bar G_k\equiv\left(\begin{array}{cc}
s_k^{\star}&-1\\
1-s_ks^{\star}_k&s_k
\end{array}\right)
\ee
One can trivially see that ${\cal L}_k\bar {\cal L}_k=\bar {\cal L}_k
{\cal L}_k=1$, and, therefore, one obtains
${\cal L}\bar {\cal L}=\bar {\cal L} {\cal L}=1$ (cf. (\ref{LM=1})).

{}From these formulas, one obtains that
$\det {\cal L}=\det \bar {\cal L}=1$ which reminds once more of the $sl(N)$
algebra. More generally, the factorization property of the
Lax operators opens the wide road to the
group theory interpretation of the RTC molecule -- see [6].

\bigskip

\noindent
{\Large\bf 6 Discrete evolutions and limit to Toda chain}

\smallskip

\noindent
{\bf Darboux-B\"acklund transformations.}
Now we are going to discuss some discrete evolutions of the RTC
given by the Darboux-B\"acklund transformations and their limit to the usual
Toda chain. One can easily take the continuum limit of the formulas
of this section to reproduce the TC as the limit of the RTC, both with
the standard continuous evolutions.

The discrete evolution equations in the RTC framework were recently
introduced by [3] in a little bit sophisticated way. Here we outline the
simple approach based on the notion of the Darboux-B\"acklund
transformation (DBT). More
details will be presented in the separate publication [16].

Let discrete index $\,i\,$ denote the successive DBT's.
The spectral problem now can be written as follows:
\be\label{Rsp}
\Phi_{n+1}(i|z) +a_n(i)\Phi_n(i|z) = z\Bigl\{ \Phi_n(i|z) +
b_n(i)\Phi_{n-1}(i|z)\Bigr\}
\ee

Let us define the first forward DBT (treating it as a
discrete analog of (\ref{todaev})):
\be \label{Rf1}
\Phi_n(i+1|z) = \Phi_n(i|z) + \balpha^{(1)}_n(i) \Phi_{n-1}(i|z)
\ee
where $\,\balpha^{(1)}_n(i)\,$ are some unknown functions.
One requires that $\;\Phi_n(i+1)\;$ satisfies the same spectral problem as
(\ref{Rsp})
but with the shifted value of $\,i\,\to i-1$.
Then the compatibility condition gives the equations of the discrete RTC:
\be\label{fa1}
a_n(i+1) = a_{n-1}(i)\;\frac{a_n(i)-\balpha^{(1)}_{n+1}(i)}{a_{n-1}(i)-
\balpha^{(1)}_{n}(i)};
\ \ \ \ \
b_n(i+1) = b_{n-1}(i)\;\frac{b_n(i)-\balpha^{(1)}_{n}(i)}{b_{n-1}(i)-
\balpha^{(1)}_{n-1}(i)}
\ee
\be\label{fz1}
z_i\frac{b_n(i)}{\balpha^{(1)}_n(i)} =
z_i - a_n(i) + \balpha^{(1)}_{n+1}(i)
\ee
where $\;z_i\;$ are arbitrary constants.

One can also introduce the discrete analog of (\ref{todaev2}):
\be\label{Rf2}
\Phi_n(i+1|z) = \bigl(1-\balpha^{(2)}_n(i)\bigr)\Phi_n(i|z) +
z\balpha^{(2)}_n(i)\Phi_{n-1}(i|z)
\ee
where $\balpha^{(2)}_n(i)\,$ are some new unknown functions of the
corresponding discrete indices. We refer to this evolution as to the
second forward DBT.
Substitution of (\ref{Rf2}) to (\ref{Rsp}) gives quite different system of the
discrete evolution equations:
\be\label{fa2'}
a_n(i+1)=
a_n(i)\frac{1-\balpha^{(2)}_{n+1}(i)}{1-\balpha^{(2)}_{n}(i)});
\ \ \ \ \
b_{n}(i+1) =
b_{n-1}(i)\frac{\balpha^{(2)}_n(i)}{\balpha^{(2)}_{n-1}(i)}
\ee
\be \label{fz2'}
a_n(i) + b_n(i)\;\frac{1-\balpha^{(2)}_n(i)}{\balpha^{(2)}_n(i)} =
z_i\;\frac{1}{1-\balpha^{(2)}_{n+1}(i)}
\ee
This system is the discrete counterpart of the continuum system (\ref{kuz1}).

Actually, in [3], four different discrete systems of the RTC equations
were written.
{}From our point of view, the additional evolutional systems result from the
{\it
backward Darboux-B\"acklund transformations} which are complimentary to those
described
above [6].

{\bf Continuum limit.}
Introducing some discrete shift of time  $\;\Delta\;>\;0\;$, one can rewrite
all the equations describing the first forward DBT as follows:
\be\label{fa11}
a_n(t+\Delta) = a_{n-1}(t)\;\frac{a_n(t)-\balpha^{(1)}_{n+1}(t)}
{a_{n-1}(t)-\balpha^{(1)}_{n}(t)}
\ee
\be\label{fb11}
b_n(t+\Delta) = b_{n-1}(t)\;\frac{b_n(t)-\balpha^{(1)}_{n}(t)}
{b_{n-1}(t)-\balpha^{(1)}_{n-1}(t)}
\ee
After the rescaling $z_i \to g\;\Delta$, one gets from (\ref{fz1})
$\balpha^{(1)}_n\sim -g\Delta b_n/a_n$
and, in the continuum limit, $\,\Delta \to 0\,$ the last two equations
lead directly to (\ref{a-eq}).

The analogous equations can be written for the second forward DBT
but with $\;z_i\;$ rescaled as $z_i \to \frac{1}{g\;\Delta}$.
It is clear that, in the limit $\,\Delta \to 0\,$, one reproduces
the continuum equations (\ref{kuz1}).

{\bf Limit to Toda chain.}
Now let us make the following expansion
(compare with (\ref{a-rep}))
\be\label{Rlim1}
a_n(i) \simeq 1-\ep p_n(i)\;\;;\;\;\;\;
b_n(i) \simeq  - \ep^2R_n(i)\\
z \simeq 1+\ep\lambda\;\; ;\;\;\;\;  z_i \simeq 1+\ep\lambda_i
\ee
Introduce also functions $\Psi_n(i)$
\be\label{Rlim2}
\Phi_n(i) \simeq \ep^n\Psi_n(i)
\ee
It is easy to see that (\ref{Rf1}) leads to the forward DBT for
the non-relativistic Toda chain if one identifies
\be
\balpha^{(1)}_n(i) \simeq \ep A_n(i)
\ee
Indeed, in the limit $\epsilon\to 0$, one gets the standard  Toda spectral and
evolution equations.

There also exist some other interesting limits [6] leading to the
modified discrete Toda equations [3].

\bigskip

\noindent
{\Large\bf Concluding remarks}

\smallskip

\noindent
From the point
of view of studying the RTC hierarchy itself, the most promising representation
is
that describing the relativistic Toda hierarchy as a particular
reduction
of the two-dimensional Toda lattice hierarchy. However, even this quite large
enveloping hierarchy is still insufficient. Loosly speaking,
the Toda lattice is too "rigid" to reproduce both the continuous and discrete
flows of the RTC.
Therefore, one should embed the RTC into a more general system which admits
more natural reductions.

This is done in the forthcoming publication [16], where we
show that the RTC has a nice interpretation if considering it
as a simple reduction of the two-component KP (Toda) hierarchy.
It turns out that, in the framework
of the 2-component hierarchy, the continuous AKNS system, Toda chain hierarchy
{\it and}
the discrete AKNS (which is equivalent to the RTC how we proved in this paper)
can be treated on equal footing.

We acknowledge V.Fock, A.Marshakov and A.Zabrodin for the discussions.
A.Z. is also grateful to V.Spiridonov, S.Suslov and L.Vinet for stimulating
discussions. The work of S.K. is partially supported by grants
RFFI-96-02-19085, INTAS-93-1038 and by  Volkswagen Stiftung, that of A.M. --
by grants RFFI-96-02-16347(a), INTAS-96-2058 and Volkswagen
Stiftung.

\bigskip

\noindent
{\Large\bf References}

\begin{itemize}

\item[1.] S.N.Ruijsenaars, {\sl Comm.Math.Phys.,} {\bf 133} (1990) 217-247

\item[2.] M.Bruschi and O.Ragnisco, {\bf A129} (1988) 21-25;
 {\sl Phys.\-Lett.,} {\bf A134} (1989) 365-370;
  {\sl Inverse Problems,} {\bf 5} (1989) 389-405

\item[3.] Yu.Suris,
 {\sl Phys.Lett.,} {\bf A145} (1990) 113-119;
{\sl Phys.Lett.,} {\bf A156} (1991) 467-474;
 {\sl  Phys. Lett.,} {\bf A180} (1993) 419-433;
 {\bf solv-int/9510007}

\item[4.] Y.Ohta, K.Kajiwara,
J.Matsukidaira and J.Satsuma,  {\bf solv-int/9304002}

\item[5.] K.Ueno and K.Takasaki,
{\sl Adv.Stud. in Pure Math.,} {\bf 4} (1984) 1-95

\item[6.] S.Kharchev, A.Mironov and A.Zhedanov, {\bf hep-th/9606144}

\item[7.] L.Faddeev and L.Takhtadjan, {\it Hamiltonian methods
 in the theory of solitons}, Sprin\-ger, Berlin, 1987

\item[8.] V.Periwal and D.Shevits,
{\sl Phys.\-Rev.\-Lett.,} {\bf 64} (1990) 1326-1335\\
M.J.Bowick, A.Morozov and D.Shevits,
 {\sl Nucl.Phys.,} {\bf B354} (1991) 496-530

\item[9.] M.J.Ablowitz and J.E.Ladik, {\sl J.Math.Phys.,}
 {\bf 17} (1976) 1011-1018

\item[10.] I.Merola, O.Ragnisco and Tu Gui-Zhang,
{\bf solv-int/9401005}\\
 A.Kundu and O.Ragnisco,
 {\bf hep-th/9401066}

\item[11.] S.Kharchev and A.Mironov,
{\sl Int.J.Mod.Phys.,} {\bf A17} (1992) 4803-4824

\item[12.] S.Kharchev, A.Marshakov, A.Mironov and A.Morozov,
{\sl Nucl.Phys.,} {\bf B397} (1993) 339-378; {\bf
hep-th/9203043}

\item[13.] R.Hirota,
{\sl Journ. of Phys.Soc. Japan,} {\bf 57} (1987) 4285-4288\\
R.Hirota, Y.Ohta and J.Satsuma,
{\sl Progr.Theor.Phys.Suppl.,} {\bf 94} (1988) 59-72

\item[14.] A.Leznov and M.Saveliev, {\sl Physica,} {\bf 3D} (1981) 62

\item[15.] M.Olshanetsky and A.Perelomov,
{\sl Phys.Rept.,}
{\bf 71} (1981) 71-313

\item[16.] S.Kharchev, A.Mironov and A.Zhedanov,
{\it Discrete and Continuous Relativistic Toda in the Framework of
2-component Toda hierarchy}, to appear

\end{itemize}

\end{document}